\documentclass[aps,amsmath,prd,preprint,floats,epsf,superscriptaddress,nofootinbib]{revtex4}
\usepackage{graphicx}
\usepackage{bm}
\usepackage{color}

\begin{document}

\preprint{IPMU 15-0111}
\bigskip

\title{Flavor-Changing Neutral-Current Decays in Top-Specific Variant Axion Model}

\author{Cheng-Wei Chiang}
\email[e-mail: ]{chengwei@ncu.edu.tw}
\affiliation{Center for Mathematics and Theoretical Physics and Department of Physics, 
National Central University, Taoyuan, Taiwan 32001, R.O.C.}
\affiliation{Institute of Physics, Academia Sinica, Taipei, Taiwan 11529, R.O.C.}
\affiliation{Physics Division, National Center for Theoretical Sciences, Hsinchu, Taiwan 30013, R.O.C.}
\affiliation{Kavli IPMU (WPI), UTIAS, The University of Tokyo, Kashiwa, Chiba 277-8583, Japan}

\author{Hajime Fukuda}
\email[e-mail: ]{hajime.fukuda@ipmu.jp}
\affiliation{Kavli IPMU (WPI), UTIAS, The University of Tokyo, Kashiwa, Chiba 277-8583, Japan}

\author{Michihisa Takeuchi}
\email[e-mail: ]{michihisa.takeuchi@ipmu.jp}
\affiliation{Kavli IPMU (WPI), UTIAS, The University of Tokyo, Kashiwa, Chiba 277-8583, Japan}

\author{Tsutomu T. Yanagida}
\email[e-mail: ]{tsutomu.tyanagida@ipmu.jp}
\affiliation{Kavli IPMU (WPI), UTIAS, The University of Tokyo, Kashiwa, Chiba 277-8583, Japan}

\date{\today}

\begin{abstract}
The invisible variant axion model is very attractive as it is free from the domain wall problem.  This model requires two Higgs doublets at the electroweak scale where one Higgs doublet carries a nonzero Peccei-Quinn (PQ) charge and the other is neutral under the PQ $\text U(1)$ symmetry.  We consider the most interesting and less constrained scenario of the variant axion model, where only the right-handed top quark is charged under the PQ symmetry and couples with the PQ-charged Higgs doublet.  As a result, the top quark can decay to the observed standard-model-like Higgs boson $h$ and the
charm or up quark, $t\to h~ c/u$, which is testable soon at the LHC Run-II.  Moreover, we propose a method to probe the chiral nature of the Higgs flavor-changing interaction using the angular distribution of $t \to ch$ decays if a sufficient number of such events are observed.  We also show that our model has the capacity to explain the $h\to\tau\mu$ decay reported by the CMS Collaboration, if the right-handed tau lepton also carries a PQ charge and couples to the PQ-charged Higgs boson.
\end{abstract}

\pacs{}

\maketitle

\section{Introduction}
\label{sec:intro}

The strong CP problem can be elegantly solved by the Peccei-Quinn (PQ) mechanism\,\cite{Peccei:1977hh}, where a $\text{U}(1)_{PQ}$ symmetry is employed to rotate away $\theta_{\rm QCD}$, the CP-violating phase in QCD.  Not manifest in the standard model (SM), the PQ symmetry must be broken spontaneously, thereby predicting the existence of a Nambu-Goldstone boson.  Since the PQ symmetry is anomalous, the additional light degree of freedom associated with the symmetry breaking is a massive pseudo Nambu-Goldstone boson, the axion\,\cite{Weinberg:1977ma,Wilczek:1977pj}.  Dynamics of the axion is characterized by the axion decay constant $f_a$.  The lower bound on $f_a$ is obtained from axion helioscopes and astronomical observations to be $f_a \gtrsim 10^9$\,GeV (see, for example, Ref.\,\cite{Agashe:2014kda}).  Moreover, coherent oscillations of the axion field can play the role of cold dark matter in the present Universe\,\cite{Abbott:1982af,Preskill:1982cy,Dine:1982ah}, from which one determines $f_a \sim 10^{11-12}$\,GeV\,\cite{Ade:2015xua} if axion is the dominant component of dark matter.  This nice mechanism, however, suffers from the problem of domain wall formation in the early Universe.  This is because the model has $N_{DW}=3$ discrete vacua related to the number of families.

The variant axion model introduced in Refs.~\cite{Peccei:1986pn,Krauss:1986wx} is an interesting axion model as it is free from the above-mentioned domain wall problem.  This is achieved by allowing only one right-handed quark to carry a PQ charge and thus rendering a unique vacuum $(N_{DW}=1)$\,\cite{Geng:1990dv}.  For consistency, the model requires two Higgs doublet fields, one of which is also charged under the PQ symmetry.  As a result, there is a non-trivial flavor structure in the Yukawa couplings\,\cite{Chen:2010su} that can lead to flavor-changing neutral-current (FCNC) couplings\footnote{To be pedantic, Yukawa couplings involve no current in the conventional sense, and FCNC would be a misnomer.  Nevertheless, we still use FCNC throughout the paper to emphasize the flavor-changing nature in the interactions mediated by the Higgs bosons between fermions of the same charge.} of the Higgs bosons to at least quarks.  Besides, such FCNC couplings depend on the helicity of fermions, that is not shared in common two-Higgs doublet models (2HDM's).   Therefore, it exhibits interesting flavor phenomena at low energies.

In this work, we consider such a 2HDM with the PQ symmetry and assign a nonzero charge to the right-handed top quark, thus dubbed the top-specific variant axion model.  In the model, Higgs-mediated FCNC couplings are generally present among the up-type quarks.  Taking into account the current SM-like Higgs data and $t \to c h$ branching ratio, we put constraints on the parameter space of the model.  We then investigate the possibility of observing the $t \to c h$ decay at the LHC.  We also show how the helicity nature in the FCNC's can be probed by studying the angular distribution of $t \to c h$ decay.  In view of the recent CMS observation of the $h \to \tau\mu$ decay\,\cite{Khachatryan:2015kon}, we also study the scenario where a nonzero PQ charge is assigned to the right-handed tau lepton as well, and find that the observed data can be accommodated by the model.  Moreover, such parameter space is within the probe of LHC Run-II on the $t \to c h$ decay.

This paper is organized as follows.  Sec.\,\ref{sec:Lagrangian} discusses the structure of the Higgs sector along with the FCNC couplings of the SM-like Higgs boson to fermions.  We study the current constraint on the mixing parameters in the Higgs sector in Sec.\,\ref{sec:current_bounds}.  In Sec.\,\ref{sec:quark_FCNC}, we concentrate on the FCNC couplings in the up-type quark sector and examine in detail the $t \to c h$ decay, including its branching ratio and asymmetry in distribution.  In Sec.\,\ref{sec:lepton_FCNC}, we turn our attention to the FCNC couplings in the lepton sector and study the $h\to\tau\mu$ decay.  Discussions and conclusions are given in Sec.\,\ref{sec:conclusion}.

\section{Top-Specific Variant Axion Model}
\label{sec:Lagrangian}

As a minimal setup of the variant axion model, we introduce two Higgs doublet fields $\Phi_1$ and $\Phi_2$ and a scalar field $\sigma$ with PQ charges $0$, $-1$ and $1$, respectively.  The gauge singlet scalar $\sigma$ gets a vacuum expectation value (VEV) $f_a$ and spontaneously breaks the PQ symmetry at a high energy scale.  It therefore does not play much a role at low energies.  In the quark sector, we assume that only the right-handed top quark field $t_R$ possesses a nonzero PQ charge of $-1$.  Note that we can additionally assign nonzero PQ charges to leptons as well, as they do not contribute to the number of axionic domain walls $N_{DW}$\,\cite{Geng:1990dv}.  We discuss such a possibility toward the end of this section.

Under the above PQ charge assignments, the most general renormalizable Higgs potential is, as given in Ref.\,\cite{Chen:2010su} and following the notation and convention of Ref.~\cite{Davidson:2005cw},
\begin{align}
V(\Phi_1,\Phi_2)
=&
m_{11}^2 \Phi_1^\dagger\Phi_1 + m_{22}^2 \Phi_2^\dagger\Phi_2
- (m_{12}^2 \Phi_1^\dagger\Phi_2 + \mbox{h.c.})
+ \frac{\lambda_1}{2} \left( \Phi_1^\dagger\Phi_1 \right)^2
+ \frac{\lambda_2}{2} \left( \Phi_2^\dagger\Phi_2 \right)^2\nonumber \\
& 
+ \lambda_3 \left( \Phi_1^\dagger\Phi_1 \right)\left( \Phi_2^\dagger\Phi_2 \right)
+ \lambda_4 \left( \Phi_1^\dagger\Phi_2 \right)\left( \Phi_2^\dagger\Phi_1 \right) ~,
\end{align}
where the $\sigma$ field has been integrated out.  The $m_{12}^2$ terms, as can be derived from the UV-complete Lagrangian~\cite{Chen:2010su}, softly violate the PQ symmetry.  Moreover, through a rotation of PQ symmetry, $m_{12}^2$ can be made real and positive.  All the other terms respect the PQ symmetry and their associated parameters ($m_{11}^2$, $m_{22}^2$, and $\lambda_{1,2,3,4}$) are real.

After the electroweak symmetry breaking, each $\Phi_i$ acquires a VEV $v_i$ and can be written in terms of component fields as $\Phi_i = (H^+_i, (v_i + h_i + iA_i)/\sqrt{2})^T$.  We define $\tan \beta = v_2/v_1$ as in the usual 2HDM's and $v^2_{\rm SM} = v_1^2 + v_2^2 \simeq 246$~GeV.  Rotating to the so-called Higgs basis~\cite{Lavoura:1994fv}, where only one of the doublets has a nonzero VEV, one has
\begin{align}
&
\begin{pmatrix}
\Phi_1 \\ 
\Phi_2
\end{pmatrix} 
= R_\beta
\begin{pmatrix}
\Phi^{\rm SM}\\ 
\Phi^\prime
\end{pmatrix}
~,~\mbox{with }
R_{\theta} =
\begin{pmatrix}
\cos\theta & -\sin\theta \\
\sin\theta & \cos\theta \\
\end{pmatrix} ~,
\\
&
\mbox{with }~
\Phi^{\rm SM} = 
\begin{pmatrix}
G^+ \\
(v_{\rm SM} + h^{\rm SM} + iG^0)/\sqrt{2}\\
\end{pmatrix}~,~ 
\Phi^{\prime} = 
\begin{pmatrix}
H^+ \\
(h^{\prime} + iA^0)/\sqrt{2}\\
\end{pmatrix} ~,
\end{align}
where $G^\pm$ and $G^0$ are the would-be Nambu-Goldstone bosons eaten by the $W^\pm$ and $Z$ bosons.  The pseudoscalar Higgs boson $A^0$ and charged Higgs boson $H^\pm$ are mass eigenstates with masses $m_A$ and $m_{H^+}$, respectively.  We also define the mass eigenstates of the CP-even neutral Higgs bosons as $h$ and $H$, with respective masses $m_h$ and $m_H$ ($m_h < m_H$), using the rotation angle $\alpha$ as follows:
\begin{align}
\begin{pmatrix}
H\\
h 
\end{pmatrix}
=
R_{-\alpha} 
\begin{pmatrix}
h_1 \\
h_2 \\
\end{pmatrix}
=
R_{ \beta - \alpha } 
\begin{pmatrix}
h^{\rm SM} \\
h^\prime \\
\end{pmatrix} ~.
\end{align}
Note that the light Higgs boson $h$ becomes a SM-like Higgs boson $h^{\rm SM}$ in the limit $\sin(\beta - \alpha) \to 1$, corresponding to the limit of $m_{12}^2 \to \infty$.  The couplings between $h$ and weak gauge bosons are read as
\begin{align}
g_{hVV}=\sin(\beta-\alpha) g_{hVV}^{\rm SM} ~,\ \ \ 
g_{HVV}=\cos(\beta-\alpha) g_{hVV}^{\rm SM} ~,\ \ \ {\rm and}\ \ \  g_{AVV}=0 ~,
\label{eq:HVV}
\end{align}
where $g_{hVV}^{\rm SM}$ are the couplings in the SM.  Since the charged Higgs boson also contributes to the $h\to\gamma\gamma$ decay, to be considered in Sec.\,\ref{sec:current_bounds}, we provide the triple Higgs coupling $\lambda_{hH^+H^-}$, defined by the $\lambda_{hH^+H^-}{hH^+H^-}$ interaction in the Lagrangian, here:
\begin{align}
\begin{split}
\lambda_{hH^+H^-}
&= \frac{1}{v_{\rm SM}}
\left[
(m_h^2 + 2 m_{H^+}^2 - 2 m_{A}^2) \sin(\beta- \alpha) 
\right.
\\
& \qquad\qquad \left. 
+ (m_A^2 + m_h^2) (\tan\beta - \cot \beta )\cos (\beta - \alpha)
\right] ,
\end{split}
\label{eq:HHH}
\end{align}
where $m_A^2 \equiv 2m_{12}^2 / \sin2\beta$.

We next investigate the Yukawa interactions.  Here we explicitly work out the result in the up-type quark sector because of the assumption that only the right-handed top quark has a nonzero PQ charge among all quark fields.  Nevertheless, the same can be done to the down-type quark sector and the lepton sector if necessary\,\cite{Chen:2010su}.  First, the up-type Yukawa Lagrangian is
\begin{eqnarray}
\mathcal L^u &=& -\Phi_1 \overline{u}_{R a} [Y_{u1}]_{ai} Q_{i} - \Phi_2 \overline{u}_{R 3} [Y_{u2}]_{i} Q_{i} + \text{h.c.}
\notag\\
&=& -\Phi^{\rm SM} \overline{u}_{R i} [Y_u^{SM}]_{ij} Q_{j} - \Phi^\prime \overline{u}_{R i} [Y_u^\prime]_{ij}  Q_{j} + \text{h.c.} ~,
\end{eqnarray} 
where the first line is written in the original basis and the second line in the Higgs basis, and the family indices run $a = 1,2$ and $i, j= 1,2,3$. The Yukawa coupling matrices, $Y_{u1}$ and $Y_{u2}$, take the following forms:
\begin{eqnarray}
Y_{u1}=\begin{pmatrix}
* &* &*\\
* &* &*\\
0 &0 &0 \\
\end{pmatrix}~,~~ 
Y_{u2}=\begin{pmatrix}
0 &0 &0 \\
0 &0 &0 \\
* &* &*\\
\end{pmatrix}~,
\end{eqnarray} 
where $*$ indicates a generally nonzero element.  As a result,
\begin{eqnarray}
Y_u^{\rm SM}  &=& \cos{\beta} Y_{u1} + \sin{\beta} Y_{u2}~,
\notag \\
Y_u^\prime &=& -\sin{\beta} Y_{u1} + \cos{\beta} Y_{u2} = \begin{pmatrix}
-\tan\beta &&\\
& -\tan\beta &\\
&& \cot\beta \\
\end{pmatrix} Y_u^{SM} ~.
\end{eqnarray} 
At this stage, the mass matrix
$M_u\equiv\frac{v_\text{SM}}{\sqrt{2}}Y_u^{\rm SM}$
is generally non-diagonal, and can be brought to such a form through a bi-unitary transformation $V M_u U^\dagger=\text{diag}(m_u,m_c,m_t)\equiv \frac{v_\text{SM}}{\sqrt2}Y_u^\text{diag}$, where $U$ and $V$ are two unitarity matrices.   In this basis, the other Yukawa matrix
\begin{eqnarray}
\begin{split}
Y_u^{\prime, \text{diag}}
&=  \begin{pmatrix}
-\tan\beta &&\\
& -\tan\beta &\\
&& \cot\beta \\
\end{pmatrix} Y_u^\text{diag} + (\tan\beta + \cot\beta)H_u
 Y_u^\text{diag},
\end{split}
\label{eq:Yuprime}
\end{eqnarray}
where the Hermitian matrix
\begin{align}
H_u\equiv V \begin{pmatrix}
0 &&\\
& 0 &\\
&&  1  \\
\end{pmatrix} V^\dagger - \begin{pmatrix}
0 &&\\
& 0 &\\
&&  1  \\
\end{pmatrix} ~.
\end{align}
Note that in the second term of Eq.~(\ref{eq:Yuprime}), the $(\tan\beta + \cot\beta)H_u$ part describes the mixing among up-type quarks and $Y_u^\text{diag}$ controls the strength of coupling with the dominant component given by the top Yukawa coupling.  From simplicity, we will omit the superscript ``diag'' while working in the mass-diagonal basis from now on.  Note that $V$ is the rotation matrix for the right-handed quarks and is completely independent of the CKM matrix, which is the product of left-handed up-type and down-type quark rotation matrices.  Therefore, the mixing angles in $V$ can be as large as $\mathcal O(1)$, a key feature of the model.

In the following, we focus exclusively on the Yukawa interactions of the observed Higgs boson, $h$.  In the mass eigenbasis,
\begin{align}
\label{eq:Yukawa}
\mathcal L_Y &\equiv 
- \sum_{f=e,\cdots,u,\cdots,d,\cdots} \xi_f \frac{m_f}{v_\text{SM}} h \overline{f}f + \mathcal{L}_\text{FCNC}
\\
& \mbox{with }~
\label{eq:Yukawa_FCNC}
\mathcal L_\text{FCNC} = 
- a \sum_{f,f'=u,c,t}(H_u)_{ff'}\frac{m_{f'}}{v_\text{SM}}h \overline{f}_Rf'_L 
+ \text{h.c.} ~,
\end{align}
where 
\begin{eqnarray}
\xi_f &\equiv&
\begin{cases}
\,\sin(\beta - \alpha) + \cot\beta \cos(\beta - \alpha) & (\mbox{for } f=t)\\
\,\sin(\beta - \alpha) - \tan\beta \cos(\beta - \alpha) & (\mbox{for } f\ne t) 
\end{cases}
\label{eq:xi} \\
a &\equiv&(\tan\beta +\cot\beta) \cos(\beta - \alpha) ~.
\label{eq:a}
\end{eqnarray}
We note in passing that there is a freedom to make the right-handed tau lepton like the right-handed top quark or the bottom quark; {\it i.e.}, $\xi_\tau = \xi_t$ or $\xi_b$.  It should be emphasized that $\mathcal L_\text{FCNC}$ in Eq.~(\ref{eq:Yukawa_FCNC}) contains not only FCNC terms, but also flavor-diagonal Yukawa interactions so that the corresponding up-type quark Yukawa couplings get modified from those in Eq.\,(\ref{eq:Yukawa}) with nonvanishing $a$ and $H_u$.  Obviously, the flavor-violating effects vanish in the limit of $\cos(\beta-\alpha) =0$.

One striking feature of $\mathcal L_\text{FCNC}$ is that flavor violation is associated with large asymmetries in chirality.  One can see this as follows: the $h\bar{f}_R f'_L$ coupling is expressed as $(H_u)_{ff'} \frac{m_{f'}}{v_\text{SM}}$ while the $h\bar{f'}_R f_L$ coupling is $(H_u)_{f'f} \frac{m_f}{v_\text{SM}}$.
Since $H_u$ is Hermitian, we have $(H_u)_{ff'} = (H_u)_{f'f}^*$.  Therefore, there is a disparity between the two couplings according to the mass hierarchy in $f$ and $f'$.  In other words, the two chiral states of the fermion $f$ have different Yukawa couplings strength with the fermion $f'$ in the opposite chiralities.

As an illustration and in anticipation of interesting collider phenomenology associated with the top quark, we restrict ourselves to the case of $t$-$c$ mixing in this paper.  The extension to general three-generation mixing is straightforward.  In such a simplified scenario, the mixing matrix $H_u$ can be parameterized as:
\begin{eqnarray}
\label{eq:mixing_parametrization}
H_u=
\frac{1}{2}\begin{pmatrix}
0 & 0 & 0 \cr
0 & 1 - \cos\rho  &   \sin\rho (\cos \phi -i\sin\phi ) \\
0 &\sin\rho (\cos \phi + i\sin\phi ) & \cos\rho -1  \\
\end{pmatrix} ~,
\end{eqnarray}
where the mixing angle $\rho$ is defined by
\[
V = \begin{pmatrix}
1 & 0 & 0 \\
0 & \cos\frac\rho2 & \sin\frac\rho2  e^{-i\phi} \\
0 & -\sin\frac\rho2 e^{i\phi} & \cos\frac\rho2
\end{pmatrix} ~.
\]
In this paper we do not consider any CP-violating effects and set $\phi$ to zero for simplicity.  Therefore, the flavor mixing phenomena are described solely by the parameter $\rho$.  The relevant FCNC terms in Eq.\,(\ref{eq:Yukawa_FCNC}) are 
$$
\mathcal L_{tc} =
-\frac{a}{2v_\text{SM}}h
\begin{pmatrix}
\bar{c}_R & \bar{t}_R
\end{pmatrix}
\begin{pmatrix}
m_c(1 - \cos\rho ) & m_t\sin\rho \\
m_c\sin\rho  & m_t( \cos\rho -1 ) \\
\end{pmatrix}
\begin{pmatrix}
c_L \cr t_L
\end{pmatrix} + \text{h.c.}\,.
$$
After redefining $\xi_f$ to include the diagonal terms in the above expression, Eq.~(\ref{eq:xi}) becomes
\begin{eqnarray}
\xi_f=\begin{cases}
\,\sin(\beta - \alpha) + \left(\cot\beta - \frac{1-\cos\rho}{2}(\tan\beta + \cot\beta)\right)\cos(\beta-\alpha) & (\mbox{for } f=t)
~, \\
\,\sin(\beta-\alpha) - \left(\tan\beta - \frac{1-\cos\rho}{2}(\tan\beta + \cot\beta)\right)\cos(\beta-\alpha) & 
(\mbox{for } f=c)
~,\\
\,\sin(\beta - \alpha) - \tan\beta \cos(\beta - \alpha) & (\mbox{for the others})~.
\end{cases} 
\label{eq:xi2}
\end{eqnarray}
These results reduce to Eq.\,(\ref{eq:xi}) by taking the mixing angle $\rho=0$.
\bigskip

So far, we have assumed that the lepton sector is not charged under the PQ symmetry, as for the down-type quark sector.  As alluded to earlier, we have the freedom to assign nonzero PQ charges to leptons without causing any problem.  As an example, we consider the scenario where only the right-handed tau lepton $\tau_R$ among the leptons also carries a PQ charge of $+1$.  Then we expect a flavor structure similar to the up-type quark sector, and $\mathcal L_\text{FCNC}$ now should also include the leptonic part:
\begin{eqnarray}
\mathcal L_\text{FCNC}^{\ell}&\equiv&  
-a\sum_{f,f'=e,\mu,\tau}(H_\ell)_{ff'}\frac{m_{f'}}{v_\text{SM}} h \overline{f}_Rf'_L  + \text{h.c.} ~,
\end{eqnarray}
where $H_\ell$ is the counterpart of $H_u$.  Clearly, this Lagrangian describes lepton flavor violation with the chirality asymmetry as in the up-type quark sector.  One can use the same parametrization as in Eq.\,(\ref{eq:mixing_parametrization}), with the mixing matrix $V^\ell$ in the lepton sector rotating the right-handed $\tau$ and $\mu$.  Note that the rotation angle $\rho_{\tau}$ can be different from $\rho$ in general.  We consider only the CP-conserving case ({\it i.e.}, the corresponding leptonic CP-violating phase $\phi_{\tau}=0$) in this paper as well.  Eq.\,(\ref{eq:xi2}) can be correspondingly tailored for the leptons and have $\xi_\tau = \xi_t$ and $\xi_\mu = \xi_c$ with the replacement $\rho\to\rho_\tau$.

\section{current Higgs data and constraints}
\label{sec:current_bounds}

In this section, we use the latest LHC Higgs data to constrain the model parameters, the angles $\alpha$, $\beta$ and $\rho$ in particular.  As noted earlier, the couplings between the SM-like Higgs boson $h$ and the SM particles are modified from their SM values: Eq.~(\ref{eq:HVV}) for the gauge bosons and Eq.~(\ref{eq:xi2}) for the fermions.  We use them to estimate the signal strengths of various Higgs production channels.

In our global $\chi^2$ fit, we take into account the signal strengths listed in Table~\ref{tab:sigs}, as reported by the ATLAS and CMS Collaborations in Refs.~\cite{ATLAS2015,Khachatryan:2014jba}.  When there are asymmetric errors, we take their average for the $\chi^2$ function.

\begin{table}[th]
\begin{tabular}{ccc}
\hline\hline
~Observable~ & ~ATLAS~\cite{ATLAS2015}~ & ~CMS~\cite{Khachatryan:2014jba}~ \\
\hline
$\mu^{\rm GGF}_{ZZ}$            & $1.7^{+0.5}_{-0.4}$      & $0.883^{+0.336}_{-0.272}$   \\
$\mu^{\rm GGF}_{WW}$            & $0.98^{+0.29}_{-0.26}$   & $0.766^{+0.228}_{-0.205}$   \\
$\mu^{\rm VBF}_{WW}$            & $1.28^{+0.55}_{-0.47}$   & $0.623^{+0.593}_{-0.479}$   \\
$\mu^{\rm GGF}_{\gamma\gamma}$  & $1.32 \pm 0.38$          & $1.007^{+0.293}_{-0.259}$   \\
$\mu^{\rm VH}_{bb}$             & $0.52 \pm 0.40$          & $1.008^{+0.527}_{-0.499}$   \\
$\mu^{\rm GGF}_{\tau\tau}$      & $2.0^{+1.5}_{-1.2}$      & $0.843^{+0.423}_{-0.382}$   \\
$\mu^{\rm VBF}_{\tau\tau}$      & $1.24^{+0.59}_{-0.54}$   & $0.948^{+0.431}_{-0.379}$   \\
\hline\hline
\end{tabular}
\caption{Signal strengths of various modes measured at LHC~\cite{ATLAS2015,Khachatryan:2014jba}.  In the first column, the superscript GGF, VBF, or VH refers to the production mechanism gluon-gluon fusion, vector boson fusion, or associated production, respectively, and the subscript indicates the channel.}
\label{tab:sigs}
\end{table}

Among the channels listed in Table~\ref{tab:sigs}, the diphoton channel is the only mode sensitive to heavy Higgs boson masses though the loop contribution of the charged Higgs boson.  However, in comparison with the SM top and $W$ loop contributions, the charged Higgs loop is less important because the scalar loop function is generally smaller.  For example, taking $M \equiv m_A = m_{H^+} = m_{H}$ in Eq.~(\ref{eq:HHH}), we find that the result is virtually independent of the choices of $M$ when it is above 200~GeV.  It is also possible to arrange a significant cancellation in the coefficient of the $\sin(\beta - \alpha)$ term in Eq.~(\ref{eq:HHH}).  We have checked in a reasonable parameter space that adding the charged Higgs contribution does not modify our final results much.  Therefore, we show in the following the results for $\lambda_{hH^+H^-}=0$.

\begin{figure}[th]
\includegraphics[width=0.45\linewidth]{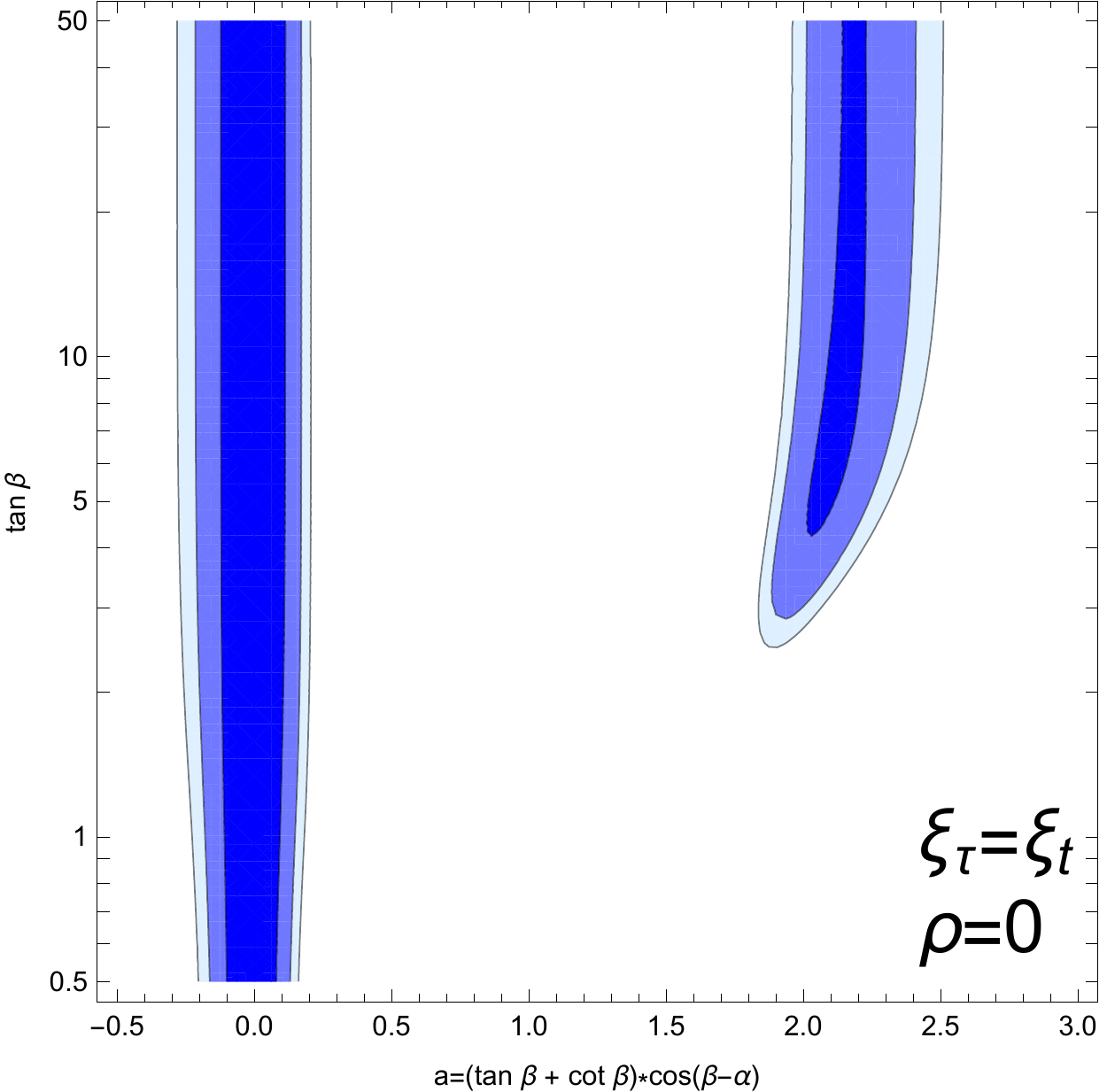}
\caption{Allowed parameter space at $68\%$ (dark blue), $95\%$ (blue) and $99\%$ (light blue) CL on the $a$-$\tan\beta$ plane.  Here we assume $\xi_\tau=\xi_t$ and $\xi_\mu=\xi_c$, and the mixing angle $\rho = 0$.}
\label{fig:atanbeta}
\end{figure}

Fig.~\ref{fig:atanbeta} shows the allowed parameter space on the $a$-$\tan \beta$ plane of our model, as constrained by the current 125-GeV Higgs data when the flavor mixing effect is switched off ($\rho=0$).  There are two branches of allowed space: $a\sim 0$ and $a\sim 2.2$.  In either branch, the parameter region has little dependence on $\tan\beta$ once it become sufficiently large.  In fact, this independence does not only occur to the $\rho = 0$ case, but also manifests when $\rho$ is finite.   At $95\%$ confidence level (CL), the branch of $a\sim 0$ is constrained to have $|a| \alt 0.2$.  The other branch corresponds to the so-called wrong-sign Yukawa region\,\cite{Ferreira:2014naa}, where the Yukawa couplings of quarks other than the top have an opposite sign to their SM ones.  In the following, we will assume large $\tan\beta\gtrsim 10$ and consider both branches.

\begin{figure}[h]
\includegraphics[width=0.45\linewidth]{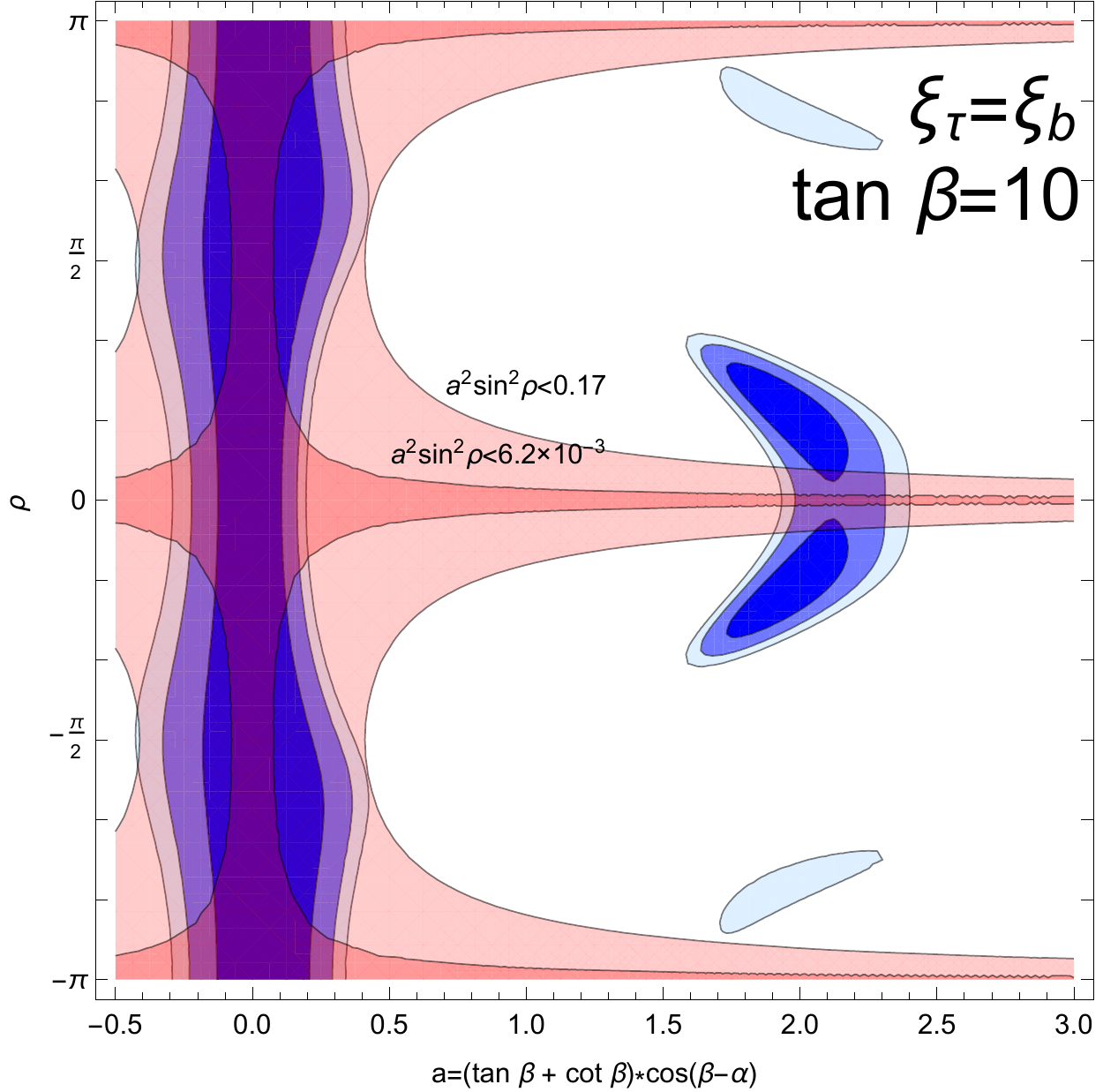}
\includegraphics[width=0.45\linewidth]{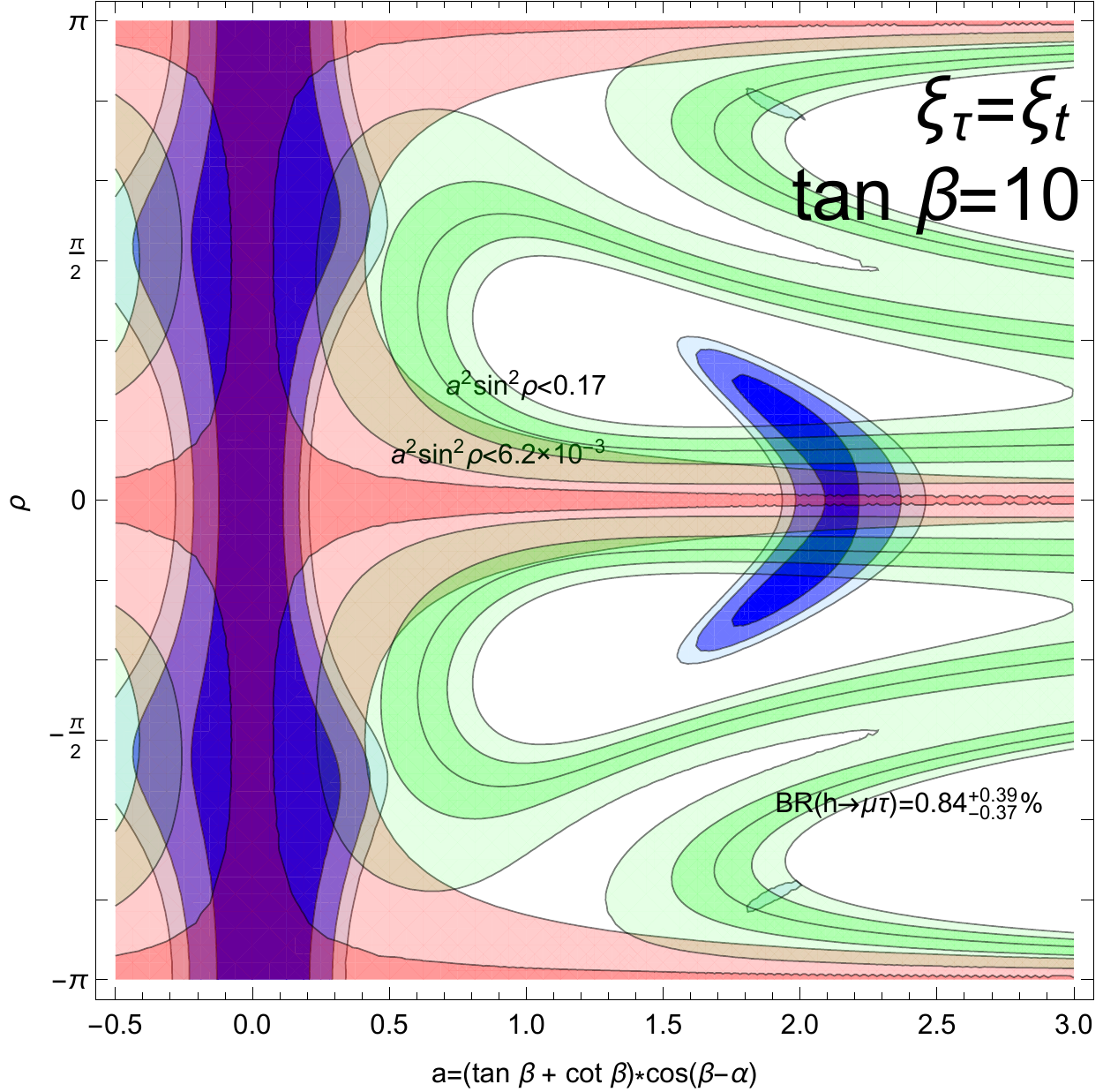}
\caption{Allowed parameter space at $68\%$ (dark blue), $95\%$ (blue) and $99\%$ (light blue) CL on the $a$-$\rho$ plane.  The left (right) plot is drawn for the $\tau$ lepton with a Yukawa coupling scaling factor $\xi_\tau$ as the bottom (top) quark.  Constraints from the current $t \to ch$ branching ratio bound at $95\%$ CL are also superimposed in pink.  The future $95\%$ CL sensitivity on the $t \to ch$ branching ratio is drawn in red.  In the right plot, the parameter regions consistent with the CMS $h \to \tau\mu$ at $68\%$ (green) and $95\%$ (light green) CL are also overlaid for comparison.}
\label{fig:arho}
\end{figure}

Next, we consider finite flavor mixing effects described by a non-zero $\rho$.  As noted above, the allowed parameter space has little dependence on $\tan\beta$ when it is $\agt 10$, we therefore fix its value at $10$ in the following numerical analysis.  Fig.~\ref{fig:arho} shows the allowed parameter space on the $a$-$\rho$ plane.  Here, we consider two scenarios: $\xi_\tau = \xi_b$ (left plot) and $\xi_\tau = \xi_t$ (right plot), corresponding to the PQ charge of $\tau$ lepton being 0 and 1, respectively.  Moreover, the mixing angle $\rho_{\tau}=\rho$ is assumed in the latter scenario.  As is expected from Fig.\,\ref{fig:atanbeta}, we find two allowed regions: $a \sim 0$ and $a \sim 2$.  The former corresponds to the decoupling limit, and the latter to the wrong-sign Yukawa limit.  A common feature in both plots is that the allowed range for $a \sim 0$ is more stringent when $\rho \sim 0, \pm \pi$ and more relaxed when $\rho \sim \pm \pi/2$.  

The region $a \sim 2$ has $\cos\rho \agt 0.85$ in both plots, with the $\xi_\tau = \xi_t$ scenario having a slightly larger allowed space.  Moreover, the scaling factor of the bottom Yukawa coupling $\xi_{b} \sim -1.2$, having an opposite sign from its SM value.  This sign change results in an increase of $\sim 30\%$ in the GGF production cross section due to the interference between the top loop and the bottom loop appearing in the effective $ggh$ coupling, which cancels with the decrease of the branching ratios due to the increase of the total width to maintain $\mu^{\rm GGF}$'s almost unchanged. Since currently the most constraining data come from the GGF production channels, we find it particularly important to measure the signal strengths $\mu_{\tau\tau}^{\rm VBF}$ and $\mu_{bb}^{\rm VH}$ to a better precision in order to probe this large-$a$ region.

As will be discussed in detail in the next section, data of FCNC processes can put useful constraints on the parameter space as well.  Such FCNC effects are proportional to $a^2 \sin^2 \rho$.  In Fig.~\ref{fig:arho}, we overlay the regions allowed by current (pink) and future (red) top FCNC measurements.  As one can see, a large portion of the large-$a$ region is excluded by such data already.  The anticipated sensitivity of the 14-TeV LHC with $3000$~fb$^{-1}$ will further constrain the large-$a$ region as well as the $a \sim 0$ region with $\rho \not= 0, \pm\pi$.

Finally, also shown in the right plot of Fig.~\ref{fig:arho} are the parameter regions consistent with $\text{BR}(h\to \tau\mu)$ reported by the CMS Collaboration, the details of which will be discussed in Section~\ref{sec:lepton_FCNC}.  
The parameter space consistent with the Higgs data and the $h\to \tau\mu$ excess at the $2\sigma$ level is found in the regions with $|a| \sim \pm 0.3$ and $\rho \sim \pm \pi/2$ and in part of the large-$a$ region.
Interestingly, these regions are only marginal in view of the current top FCNC bound and within the reach of LHC Run-II experiment.

\section{Top quark FCNC interactions}
\label{sec:quark_FCNC}

This model generically predicts the top FCNC decay $t \to ch$ (or $t \to uh$) via the mixing, serving as a smoking gun signature.  
For definiteness, we only refer to $t \to ch$ in this section, but note that basically current experimental limits do 
not tag the flavor of the accompanied jet and what is constrained is the sum of all branching ratios of $t \to qh$.

At the LHC, one can use photon channel and multi-lepton channels to search for $t \to ch$ in top pair production, {\it i.e.}, $pp \to t\bar{t} \to (b\ell\nu)(ch)$ with $h \to WW/ZZ/\tau\bar{\tau}$~\cite{Craig:2012vj}. Both ATLAS using $h \to \gamma\gamma$ channel (7-TeV and 8-TeV)~\cite{Aad:2014dya} 
and CMS through inclusive multi-lepton channels  (8-TeV with 19.5~fb$^{-1}$)~\cite{Chatrchyan:2014aea} searched for $t\to ch$ and set the upper bounds on the branching ratio of 0.79\% and 1.3\%, respectively, which correspond to $\sqrt{|\lambda_{tc}|^2 + |\lambda_{ct}|^2}  < 0.17$ and $< 0.21$, where $\lambda_{tc}$ and $\lambda_{ct}$ are the flavor-changing Yukawa coupling appearing in the interaction Lagrangian, $- (\lambda_{tc} \bar{t}_R c_L + \lambda_{ct} \bar{c}_R t_L)h +$ h.c.
CMS further improved its limit by adding the leptons + di-photon channel in the same event set as above,
and obtained the constraint $\text{BR}(t \to ch) < 0.56\%$ at 95\% CL, corresponding to $\sqrt{|\lambda_{tc}|^2 + |\lambda_{ct}|^2} < 0.14$~\cite{Khachatryan:2014jya,CMS-PAS-HIG-13-034}.

In the future LHC run, the limits on $\text{BR}(t \to ch)$ will be greatly improved, with the expectation of $2 \times 10^{-3}$ (300~fb$^{-1})$ or $5 \times 10^{-4}$ (3000~fb$^{-1})$ in the lepton channels and $5 \times 10^{-4}$ (300~fb$^{-1})$ or $2 \times 10^{-4}$ (3000~fb$^{-1})$ in the photon channels~\cite{ATL-PHYS-PUB-2013-012, Agashe:2013hma, ATLAS:2013hta}.  Note that these numbers are estimated simply by scaling up the cross sections without optimizations of the analysis.  Therefore, better bounds in reality are anticipated.  A more optimistic study~\cite{AguilarSaavedra:2004wm} predicts a branching ratio of $5.8 \times 10^{-5}$ as the $3\sigma$ discovery limits with 100~fb$^{-1}$ using the $h \to b\bar{b}$ mode.  One can also probe the same coupling using the single top and $h$ associated production: $ug \to th$ and $cg \to th$.  However, the sensitivity of such processes at the LHC is usually weaker than the rare top decay search.

It is noted that the $h$-$t$-$c/u$ coupling can also contribute to other flavor observables.  For example, imaginary parts of the flavor-violating Yukawa couplings are constrained by the hadronic electric dipole moments, CP-violating observables in the $D$-meson sector, and $D$-meson mixing~\cite{Gorbahn:2014sha}.

\subsection{Signal rate}

Using the parameters defined in Section~\ref{sec:Lagrangian}, one can compute the decay amplitudes ${\cal M}_{h_t,h_c}$ as follows:
\begin{eqnarray}
{\cal M}_{+,+} = \frac{m_t a \sin\rho}{2v}\sqrt{2m_t E_3}\cos\frac{\theta}{2} ,\ \ 
{\cal M}_{-,+} = -\frac{m_t a \sin\rho}{2v}\sqrt{2m_t E_3}\sin\frac{\theta}{2},\ \ 
{\cal M}_{\pm,-}=0,
\label{topdecay}
\end{eqnarray}
where the subscripts $h_t$ and $h_c$ $(=+$ or $-)$ denote the spin direction of the top quark and the helicity of the charm quark, respectively.  The angle $\theta$ is defined as the charm momentum direction relative to the top spin direction in the top rest frame.  We have neglected the term proportional to $m_c$ compared to $m_t$ and, as a result, the amplitude involving the left-handed charm quark is vanishing.

The partial decay width of $t\to ch$ is computed as
\begin{align}
\begin{split}
\Gamma_{t \to ch}
&= \frac{G_F m_t^3  a^2  \sin^2\rho}{64 \pi \sqrt{2}}(1-r_h^2)^2,
\\
&
\mbox{with }~ E_3=\frac{m_t}{2} \left( 1 - \frac{m_h^2 - m_c^2}{m_t^2} \right) 
~\mbox{ and }~ r_h^2 \equiv \frac{m_h^2}{m_t^2} \sim 0.522 ~,
\end{split}
\end{align}
where we take $m_h=125$~GeV and $m_t=173$~GeV.  We can obtain the branching ratio by comparing it with the width of $t \to bW$ in the SM at the leading order,~\footnote{It is known that QCD corrections at next-to-leading order result in an ${\cal O}(10\%)$ reduction in the partial width of $t \to bW$~\cite{Chetyrkin:1999br}.  For consistency, however, we use the Born widths for both $t \to bW$ and $t \to ch$ to evaluate the branching ratio of the latter. Even if the QCD corrections are included, they should roughly cancel out in the branching ratio.}
assuming $\text{BR}(t \to bW)$ is close to unity:
\[
\Gamma_{t\to Wb} = \frac{G_F |V_{tb}|^2 m_t^3}{8\pi \sqrt{2}}(1-r^2_W)^2(1+2r^2_W) ~,
\]
where $r_W^2\equiv m_W^2 / m_t^2 \simeq 0.214$.  That is, 
\begin{align}
\text{BR}( t \to c h) = 
\frac{(1-r_h^2)^2}{8(1-r_W^2)^2(1+2 r_W^2) |V_{tb}|^2}a^2  \sin^2\rho \simeq
(3.24 \times  10^{-2} )  a^2  \sin^2\rho ~.
\end{align}
The current branching ratio limit of 0.56~\% constrains the mixing parameter as
\[ 
a^2 \sin^2\rho < 0.17 ~.
\]
The future sensitivity of $\text{BR}( t \to c h) < 2 \times 10^{-4}$ (14~TeV and 3000~fb$^{-1}$) will set the limit 
\[ 
a^2 \sin^2 \rho < 6.2\times 10^{-3} ~.
\]
Such constraints have been overlaid in Fig.~\ref{fig:arho} by the pink and red regions, respectively.

\subsection{Decay distribution}

Once we observe enough $t \to ch$ events, it will be possible to check the chiral property in the flavor-changing coupling as predicted in the model, namely, the charm quark in the decay product should be right-handed.  For this purpose, we can make use of the helicity amplitudes of top decay in Eq.~(\ref{topdecay}) once we know the top spin direction.  The spin analyzing power $\kappa_i$ of the decay product $i$ is defined as 
\begin{eqnarray}
\frac{1}{\Gamma_i}\frac{d\Gamma_i}{d\cos\theta_i} = \frac{1}{2}( 1 + \kappa_i P \cos\theta_i) ~,
\end{eqnarray}
where $P$ is the polarization of the decaying particle along a certain direction, $\Gamma_i$
 is the partial decay width into the decay product $i$, and 
$\theta_i$ is the polar direction of the decay product $i$ relative to 
the polarization axis. The spin analyzing power 
of the charged lepton $\kappa_{\ell^+}$ from the usual top decay 
$t \to b\ell^+\nu$ is known to have largest value $+1$ at leading 
order~\cite{Bernreuther:2008ju}.  Our model predicts $d\Gamma_{t \to ch}/d\cos\theta \propto 1+\cos\theta$.  Therefore, the charm quark and the Higgs boson have the 
spin analyzing powers $\kappa_c=+1$ and $\kappa_h=-1$, respectively.  We  denote the spin analyzing power for the anti-top with $\bar{\kappa}$ and note that $\bar{\kappa}_{\bar{f}}= -\kappa_f$ assuming CP invariance.

The net polarization of the top quark is zero in the top pair production at the LHC.  Still, we can use either $t\bar{t}$ spin correlation or single top production as the source of initial top spin polarization.  As the spin analyzing power for charm (or Higgs) has a maximal value in the model, we expect that the spin-correlation analysis using the di-lepton channel in top pair production at the LHC~\cite{Aad:2014mfk, CMS:2015dva} should work similarly using the lepton + di-photon channel from $t\bar{t} \to (b\ell^+\nu)(\bar{c}h)$ or $(\bar{b}\ell^-\bar{\nu})(ch)$.  Moreover, this channel involves only one neutrino as the source of missing momentum and, therefore, one can completely solve the event kinematics
using top and $W$ mass shell conditions.

\begin{figure}[bt]
\includegraphics[width=0.5\linewidth]{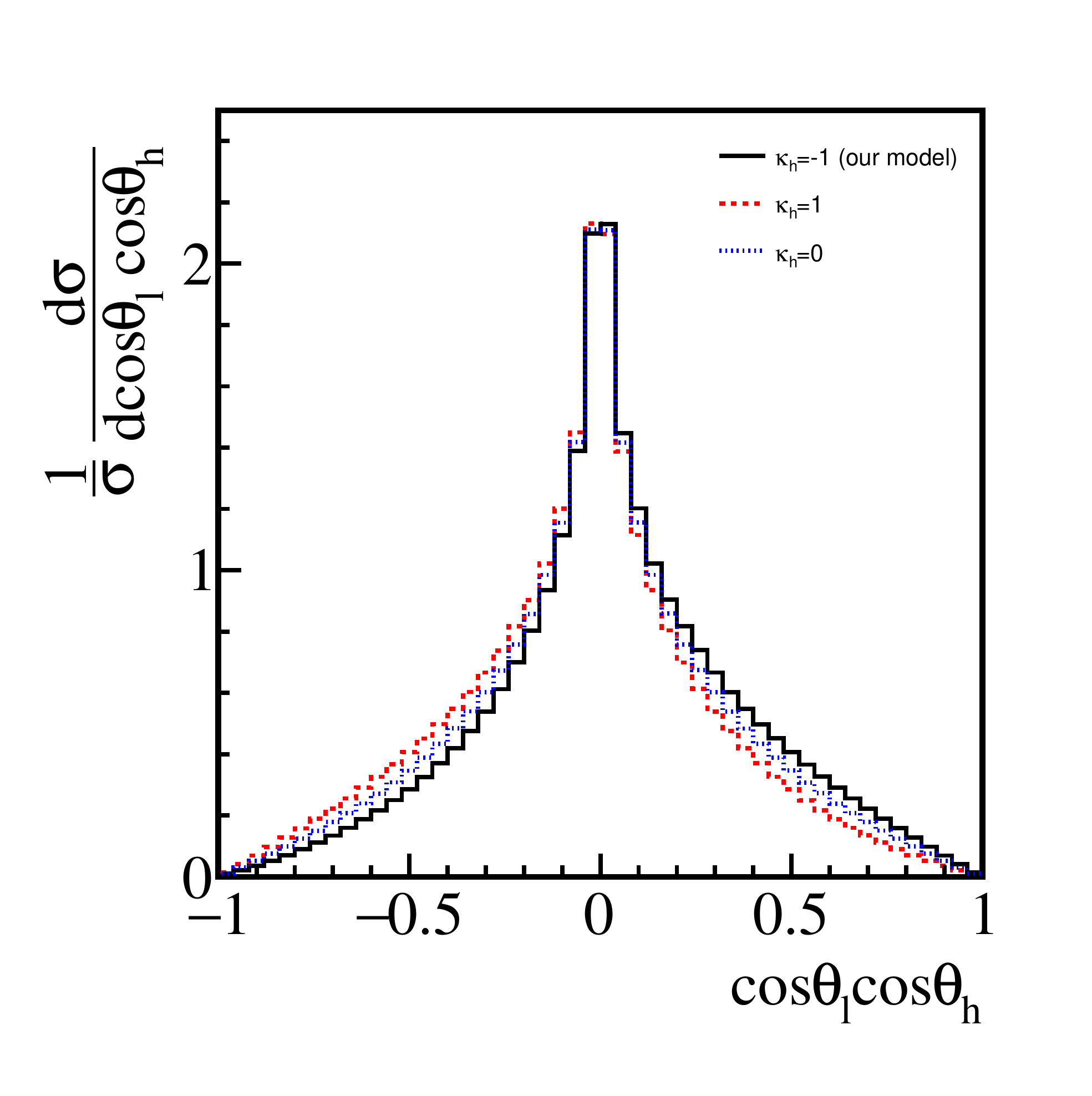}	
\caption{Expected $\cos\theta_\ell \cos\theta_h$ distributions for $\kappa_h=-1$(our model), $0$ and $+1$.
}
\label{fig:asym}
\end{figure}

At the LHC, using the helicity basis is known to be a reasonably good spin quantization axis to probe the spin correlation effect~\cite{Bernreuther:2004jv}.  The asymmetry defined by
\begin{eqnarray}
A_{\rm hel}=\frac{N(t_\uparrow \bar{t}_\uparrow) + N(t_\downarrow \bar{t}_\downarrow) - N(t_\uparrow \bar{t}_\downarrow) - N(t_\downarrow \bar{t}_\uparrow)}
{N(t_\uparrow \bar{t}_\uparrow) + N(t_\downarrow \bar{t}_\downarrow) + N(t_\uparrow \bar{t}_\downarrow) + N(t_\downarrow \bar{t}_\uparrow)}
\end{eqnarray}
is predicted to have a value $\sim 0.35$.  This asymmetry shows up in the double theta distribution:
\begin{eqnarray}
\frac{1}{\sigma} \frac{d\sigma}{d\cos\theta_i d\cos\theta_j} = \frac{1}{4}( 1 + A_{\rm hel} \, \kappa_i \bar{\kappa}_j \cos\theta_i \cos\theta_j) ~,
\end{eqnarray}
where $\theta_{i,j}$ are defined in the rest frame of top and anti-top quark, respectively.
We can determine $\kappa_h$ (or $\kappa_c$)  by measuring the distribution for 
$i=\ell^+$, $j=h$ and for the corresponding anti-particle case.
We expect that the distributions of $\cos\theta_{\ell^+}\cos\theta_h$  
and $\cos\theta_{\ell^-}\cos\theta_h$ should be identical due to the relation $\kappa_{\ell^+}\bar{\kappa}_h = \bar{\kappa}_{\ell^-}\kappa_{h}$.  Therefore, we simply use the notation $\cos\theta_\ell\cos\theta_h$ in the following.  Fig.~\ref{fig:asym} shows the expected distribution of $\cos\theta_\ell\cos\theta_h$ for $\kappa_h=\pm 1, 0$ cases. 
One can obtain the coefficient $A_{\rm hel}\, \kappa_{\ell^+} \bar{\kappa}_h$ by determining the mean value of the quantity under this distribution:
\begin{eqnarray}
\left< \cos\theta_\ell \cos\theta_h\right> = \frac{A_{\rm hel}\, \kappa_{\ell^+} \bar{\kappa}_h}{9} \end{eqnarray}
In our model, $ \kappa_h = -\bar{\kappa}_h= -1$ 
gives a positive mean value of $\sim 0.039$.

Finally, we provide a rough estimate for the sensitivity based on a simpler observable instead, and leave the detailed study to a future work.  Let's introduce the asymmetry 
\begin{eqnarray}
A_{\ell h}=\frac{N( \cos\theta_\ell \cos\theta_h >0) - N( \cos\theta_\ell \cos\theta_h <0)}{N( \cos\theta_\ell \cos\theta_h >0) + N( \cos\theta_\ell \cos\theta_h <0)}=
\frac{A_{\rm hel} \kappa_{\ell^+} \bar{\kappa}_h}{4} \sim 0.088 \bar{\kappa}_h.
\end{eqnarray}
To confirm $\kappa_h \sim -1$, we have to measure $A_{\ell h}$ with a precision better than 0.088.
The statistical uncertainty on $A_{\ell h}$ is given by 
\begin{eqnarray}
\Delta A_{\ell h} \simeq \Delta N/N \simeq 1/\sqrt{N} > 0.088 ~,
\end{eqnarray}
which implies that we need at least $130$ signal events to confirm the decay distribution structure at $1\sigma$ level.
This simple estimate does not include background estimates.  However, with $\sigma(t\bar{t}) \sim 1$~nb 
at the 14-TeV LHC and an integrated luminosity of 3000~fb$^{-1}$, we expect 
$3 \times 10^9$ top pair events. 
It provides $\sim 10^6$ $t\to ch$ events if $\text{BR}(t\to ch) =2 \times 10^{-4}$, a conservative sensitivity that LHC can reach.  Even after multiplying $\text{BR}(h\to \gamma\gamma) \sim 2.3 \times 10^{-3}$ and the leptonic decay branching ratio of the top quark, we still expect $\sim 500$ events.  Besides, the $h \to b\bar{b}$ mode can be incorporated to enhance the number of events~\cite{AguilarSaavedra:2004wm}.  

\section{Lepton FCNC interactions and $h\to\tau\mu$ decay}
\label{sec:lepton_FCNC}

As we have shown in Section\,\ref{sec:Lagrangian}, the Higgs boson can have lepton flavor-changing Yukawa couplings if the right-handed $\tau_R$ carries a nonzero PQ charge.  In this section, we discuss the lepton FCNC interactions and the corresponding constraints.  The CMS Collaboration recently reported a measurement of the flavor-violating decay $h\to\tau\mu$~\cite{Khachatryan:2015kon}.  
The observed branching ratio for this decay mode
\begin{equation}
\text{BR}_{\rm obs}(h \to \mu\tau) = \frac{N_{\rm obs}}{\mathcal{L~A}~ \sigma_\text{SM}}
= (0.84^{+0.39}_{-0.37})\text{\,\%} ~,
\label{eq:CMS_FCNC}
\end{equation}
where $N_{\rm obs}$ denotes the number of observed events, $\mathcal L$, $\mathcal A$ and $\sigma_{\rm SM}$ are the integrated luminosity, the acceptance of the selection cuts, and the total production cross section in the SM, respectively.  
To relate this quantity to our model, we have to including a correction factor in the production cross section:
\begin{equation}
\text{BR}_{\rm obs}(h \to \mu\tau) = \text{BR}_\text{VA}(h \to \mu\tau)\frac{\sigma_\text{VA}}{\sigma_\text{SM}}
\simeq {\xi_g}^2 \text{BR}_\text{VA}(h \to \mu\tau),
\label{eq:BRobs}
\end{equation}
where $\xi_g$ is the $ggh$ coupling scale factor in our model relative to the SM one
and we have assumed that the production is dominated by the GGF mechanism.
The VBF mechanism also partly contributes to the cross section.  We have checked that including the VBF production would infer a smaller value of $a$ to explain the signal.  Our result above is thus considered as a more conservative estimate.

The branching ratio in the model is found to be
\begin{align}
\text{BR}_\text{VA}(h \to \mu\tau) 
\simeq\frac{a^2 \sin^2\rho_{\tau}}{36.52{\xi_b}^2 + 14.64 \sin^2(\beta-\alpha) + 5.44{\xi_g}^2 + 4{\xi_\tau}^2} ~,
\label{eq:BRVA}
\end{align}
where we have included only the $h\to bb,WW^\star, ZZ^\star, gg$ and $\tau\tau$ decays in the denominator.
Therefore, we obtain for $\xi$'s $\simeq 1$ that
\begin{equation}
a^2 \sin^2 \rho_{\tau} \sim 0.35,
\end{equation}
in order to explain the observed $h\to\tau\mu$ events.  We show the corresponding parameter regions using the green bands in Fig.\,\ref{fig:arho}.  The green (light green) band width corresponds to the $1\sigma$ ($2\sigma$) error quoted in Eq.~(\ref{eq:CMS_FCNC}).  As seen in the plot, even with maximal mixing $|\sin\rho_{\tau}| \sim 1$ we required $a \simeq 0.6$.

This model predicts the same helicity structure as in the top sector discussed in the previous section, that is, decaying Higgs provides left-handed $\tau^-$ in $h \to \tau^- \mu^+$ decay.
We can confirm it by observing the energy fraction carried by the visible decay products of the $\tau$.
For the left-handed $\tau$ decay the visible energy fraction distribution is expected softer~\cite{Hagiwara:1989fn,Rainwater:1998kj,Papaefstathiou:2011kd}.
The fact that full event kinematics reconstruction is possible in this process makes the 
analysis straightforward.

Finally, we comment on the related lepton flavor-violating decay: $\tau\to\mu\gamma$.  It is found that, as also discussed in Ref.\,\cite{Omura:2015nja}, the decay rate of $\tau\to\mu\gamma$ is well below the present experimental upper bound in the parameter region for explaining the $h\to\tau\mu$ events.

\section{conclusion}
\label{sec:conclusion}

In this work, we have studied the top-specific variant axion model and the phenomenology related to the flavor-changing Yukawa couplings of the top quark.  This model is well-motivated to solve the strong CP domain wall problem.  Only the right-handed top quark field is charged under the Peccei-Quinn (PQ) symmetry.  Two Higgs doublet fields are required for consistency and one of them also carries a nonzero PQ charge.  Compared with the usual two-Higgs doublet models, this model has additional flavor mixing parameters.  We investigated in detail the scenario with top-charm mixing, governed by a mixing parameter $\rho$, leading to modifications of Yukawa couplings and predicting the $t \to hc$ decay.  We constrained the model parameter space using the current Higgs data on signal strengths of different channels and bounds on the branching ratio of $t \to hc$ decay.  While the decoupling limit is favored by the data, a small parameter space with the wrong sign in bottom Yukawa coupling still cannot be ruled out at the moment.  Moreover, we found that a large portion of the allowed parameter space will be covered by the $t \to hc$ search at LHC Run-II.

The helicity structure in the flavor-changing top Yukawa coupling is a specific feature of the model.  The coupling between the left-handed top and the right-handed charm is stronger than that in the other chirality combination.  We illustrated that this property could be measured in the decay distribution of spin-correlated top pair production once a sufficient number of signal events are collected.  We also note that the sensitivity assumed here is conservative and
further improvement in the sensitivity of the $t \to hc$ decay measurement at the LHC would be possible and 
very important.

We have also considered the scenario where the right-handed $\tau$ lepton also carries a nonzero PQ charge, as the right-handed top quark, and couples to the PQ-charged Higgs boson.  In this case, we also expect to have the $h\to \tau \mu$ decay.  Interestingly, the parameter region allowed by the Higgs signal strength data was slightly enlarged.  We showed that the parameter regions required by the anomalous $h \to \tau\mu$ branching ratio reported by the CMS Collaboration had parts consistent with the above-mentioned constraints.  The overlapped regions were marginal to the bounds on $\text{BR}(t \to hc)$ and well within the reach of the future LHC sensitivity.  If so, we would expect the $t \to hc$ signal 
to be soon observed at the LHC Run-II.  By measuring the tau polarization in the flavor-changing Higgs decay, we could also 
probe the specific helicity structure in the lepton Yukawa couplings of the model.

\section*{Acknowledgments}

The authors would like to thank M.~E.~Peskin for some discussions about the $t \to h c$ decay that stimulates this analysis.  This research was supported in part by the Ministry of Science and Technology of Taiwan under Grant No.\ NSC 100-2628-M-008-003-MY4 (C.-W.~C); and in part by the Grants-in-Aid for Scientific Research from the Ministry of Education, Culture, Sports, Science, and Technology (MEXT), Japan No.~26104009 and Grant-in-Aid No.~26287039 from the Japan Society for the Promotion of Science (JSPS) (T.~T.~Y.); and the World Premier International Research Center Initiative (WPI), MEXT, Japan (H.~F., M.~T. and T.~T.~Y.)..

\end{document}